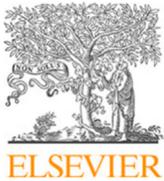
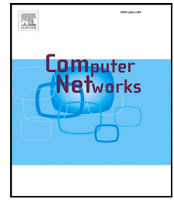

# Vehicular Cloud Computing: A cost-effective alternative to Edge Computing in 5G networks

Rosario Patanè [a],*, Nadjib Achir [c], Andrea Araldo [b], Lila Boukhatem [a]

[a] *Université Paris-Saclay, CNRS, Laboratoire Interdisciplinaire des Sciences du Numérique, 91400, Orsay, France*
[b] *Telecom SudParis, SAMOVAR, IP-Paris, France*
[c] *INRIA, France*

A B S T R A C T

Edge Computing (EC) is a computational paradigm that involves deploying resources such as CPUs and GPUs near end-users, enabling low-latency applications like augmented reality and real-time gaming. However, deploying and maintaining a vast network of EC nodes is costly, which can explain its limited deployment today. A new paradigm called Vehicular Cloud Computing (VCC) has emerged and inspired interest among researchers and industry. VCC opportunistically utilizes existing and idle vehicular computational resources for external task offloading.

This work is the first to systematically address the following question: *Can VCC replace EC for low-latency applications?* Answering this question is highly relevant for Network Operators (NOs), as VCC could eliminate costs associated with EC given that it requires no infrastructural investment. Despite its potential, no systematic study has yet explored the conditions under which VCC can effectively support low-latency applications without relying on EC. This work aims to fill that gap.

Extensive simulations allow for assessing the crucial scenario factors that determine when this EC-to-VCC substitution is feasible. Considered factors are load, vehicles mobility and density, and availability. Potential for substitution is assessed based on multiple criteria, such as latency, task completion success, and cost. Vehicle mobility is simulated in SUMO, and communication in NS3 5G-LENA. The findings show that VCC can effectively replace EC for low-latency applications, except in extreme cases when the EC is still required (latency < 16 ms).

## 1. Introduction

At the beginning of the 21st century, major data center companies realized that their computing resources were not being fully utilized during regular working hours. Despite this underutilization, these resources still consumed energy to remain operational [1]. To address this inefficiency, companies started renting out their unused computing resources through pay-on-demand models [2]. Notable examples include the introduction of Amazon Web Services (AWS) in 2006 [3] and Google App Engine in 2008 [4], both of which popularized commercial Cloud services based on the concept of Cloud Computing (CC).

Currently, the Cloud provides essential services such as storage and computing, making the concept of CC a practical and integral part of modern applications. Unfortunately, as the demand for stringent delay requirements evolves, CC is unable to meet these increasingly critical application needs. Indeed, while CC can theoretically handle large computational and storage resources [5], its latency is often excessive for modern delay-sensitive applications. For instance, according to Verizon's continuous network latency monitoring [6], typical round-trip delays are about 35 ms within North America and 70 ms for transatlantic communications. These values serve as a reliable baseline for assumptions about Cloud latency. As a result, a new paradigm emerged with the idea of geographically bringing Cloud services closer to end-users, giving rise to the concept of Edge Computing (EC) [7].

This new paradigm places computational resources closer to users, which may significantly decrease the delays [8]. Nevertheless, EC has some limitations compared to CC. Indeed, enabling EC requires deploying numerous Edge nodes across a wide geographic area. This endeavor turns out to be very costly, underscoring the need for research studies on the investment required for Edge resource deployment [9].

* Corresponding author.
  *E-mail addresses:* rosario.patane@universite-paris-saclay.fr (R. Patanè), nadjib.achir@inria.fr (N. Achir), andrea.araldo@telecom-sudparis.eu (A. Araldo), lila.boukhatem@universite-paris-saclay.fr (L. Boukhatem).

As a consequence, new "Edge" types have recently been proposed. Some of them have introduced the concept of Vehicular Cloud Computing (VCC) [10]. VCC entails creating a CC composed of the vehicle resources not used by the vehicles' On-Board Units (OBUs). VCC resources have already been deployed or are under deployment thanks to the proliferation of new connected and autonomous vehicles. For example, Tesla has equipped its cars with powerful NVIDIA GPUs [11, 12] and the embedded computational resources of these vehicles would be highly valuable in reducing the investment in Edge Computing and infrastructure costs if a sufficient number of vehicles is available to satisfy task offloading requests [13].

Our previous work, [14], investigated whether VCC can effectively replace EC when adopting Wave-IEEE 802.11p communication technology between the vehicles and infrastructure. This study identified several parameters that enable VCC as an alternative to EC (such as traffic load, vehicle density, and VCC computational resources) and showed that VCC can satisfy low-latency applications across a wide range of these parameters.

To meet modern-day real-time vehicle communication needs such as autonomous driving, 5G represents a suitable alternative to IEEE 802.11p as it brings higher performance in terms of latency, capacity, and scalability. This is further supported by industry trends and regulatory decisions, such as the Federal Communications Commission (FCC) in the USA [15], reallocating spectrum to 5G-based technologies in recognition of its potential to enhance automotive safety and connectivity.

This study is the first to investigate the conditions under which VCC can serve as a viable alternative to EC for task offloading in 5G networks. In this regard, this paper presents:

- An extensive simulation campaign in which the mobility is modeled in SUMO and the 5G communication in NS3. The impact of the following factors is analyzed: vehicle density, computational power, task workload, and vehicle mobility in urban and non-urban scenarios. The criteria considered are latency requirements and task failure rate.
- A cost analysis model that shows cost savings offered by VCC over EC, making VCC attractive for network operators willing to provide offloading services.

The rest of this paper is structured as follows. In Section 2, the paper delves into the related work, then on the system model in Section 3. Section 4 presents the adopted task offloading strategies, and details on task models are given in Section 5. Section 6 presents the performance evaluation and simulation results. Section 7 provides a preliminary analysis of cost savings due to the substitution of EC by VCC and the use of 5G. Finally, Section 8 concludes the paper.[1]

The software used in this work is open source and provided on GitHub.[2]

## 2. Related work

Task offloading allows devices, such as smartphones or OBUs, to execute tasks in external nodes. This reduces the consumption of the device's battery or to execute complex tasks that would require a large amount of computation or special hardware that may not be available in the device itself [16]. Task offloading is generally performed on the Cloud. However, some task applications, e.g., online gaming and Autonomous Driving (AD), may not tolerate the latency to reach the Cloud [17]. To execute tasks in the proximity of the device, EC has been extensively studied: by executing tasks in Edge nodes, such as Base Stations (BS) or Wi-Fi APs, the latency can be greatly reduced. VCC could contribute to real-world task offloading as an alternative to EC. A vast amount of research has been devoted to task offloading strategies when tasks are offloaded from vehicles to Edge nodes to perform low latency AD functions [18,19].

---

[1] Table 1 provides the list of acronyms.
[2] https://github.com/patan3saro/ComputationEcosystem_in_5G.git

**Table 1**
Acronyms.

| Acronym | Meaning |
|---|---|
| CC | Cloud Computing |
| EC | Edge Computing |
| MEC | Mobile Edge Computing |
| VCC | Vehicular Cloud Computing |
| AWS | Amazon Web Services |
| AD | Autonomous Driving |
| OBU | On-Board Unit |
| SUMO | Simulator of Urban MObility |
| UE | User Equipment |
| VUE | Vehicle User Equipment |
| PUE | Pedestrian User Equipment |
| BS | Base Station |
| AP | Access Point |
| CN | Core Network |
| OF | Optical Fiber |
| SGW | Serving Gateway |
| UDP | User Datagram Protocol |
| gNodeB (or gNB) | Next Generation Node B |
| FR1 | Frequency Range 1 |
| NR | New Radio |
| 3GPP | Third Generation Partnership Project |
| TTI | Transmission Time Interval |
| GPU | Graphics Processing Unit |
| MIPS | Million Instructions Per Second |
| MI | Million Instructions |
| FIFO | First In, First Out policy |
| QoE | Quality of Experience |
| ANOVA | Analysis of Variance |
| NS3 | Network Simulator version 3 |
| CAPEX | Capital Expenditure |
| OPEX | Operational Expenditure |
| AR | Augmented Reality |
| NO | Network Operator |

However, EC deployment at the level of BS and APs is still nonexistent due to its high cost of deployment [9]. This is why we examine in this work whether offloading on vehicles can replace offloading on Edge nodes.

Other works investigate solutions where vehicles share computational resources among themselves. In [20,21], the authors consider two types of vehicles: Task Vehicles (TaVs) and Service Vehicles (SeVs). When the TaVs require additional resources to complete their tasks, the SeVs make their resources available to help them complete their workloads. Although those works have shown that VCC can behave as an Edge on the road, they suffer from inherent limitations. Paper [21] demonstrated that task allocation takes numerous seconds against the low-latency application requirements [22]. Paper [20] reported lower latency in the tens of milliseconds but introduced decision overhead on vehicles, which suffer from limited communication coverage (200 m). Finally, works like [23,24], authors proposed methods for resource allocation in VCC, by integrating Edge of Cloud infrastructure with vehicular ad hoc networks. Unfortunately, these works suffer from both limited communication range and relatively high latency delays, primarily because they rely on IEEE 802.11p-based communication technologies. In our work, 5G overcomes these issues of offloading decision overhead by providing broader coverage.

In [25], the authors focus on road safety-related offloading in areas with poor Edge node coverage, resulting in limited computational resources but at the expense of non-negligible computational overhead of the adopted node selection strategy. Moreover, unlike our work, which focuses on real-world urban scenarios (using the SUMO tool), the authors considered a simple mobility model assuming constant speed on a highway. The findings of [25] do not directly apply to our work, as this paper considers EC and vehicles as complementary and used together. In contrast, our work focuses on fully replacing EC, avoiding the expenses related to its deployment and maintenance.

The literature also presents architectures composed of EC, VCC, and CC, sharing their advantages. In [26], the authors consider a three-tier model comprising VCC, EC, and CC in which the three components operate in the same hierarchy. They propose a collaborative approach for computation offloading using vehicles with idle resources as Fog User Equipment (F-UE). However, vehicle mobility and communication channels are assumed to be semi-static, whereas we consider realistic time-varying channels and user mobility.

One of the first papers to introduce the use of a three-tier approach is [13]. The authors suggest an offloading scenario where task processing requests are initiated by user mobile terminals and managed by a Cloudlet connected to a VCC through Wi-Fi APs. End devices can connect to the Cloudlet via a BS using 3G/4G connections or directly through APs. The Cloudlet, which serves as the control center, determines whether a task should be offloaded to the Cloud, the Cloudlet itself, or a VCC. Inspired by this work, we adopt a similar architectural model, as it is described in Section 3.

Recent research, such as in [27], also relied on an integrated CC, EC, and VCC architecture. The authors aimed to minimize task offloading time for end-users by exploiting the collaborative effort of the three component tiers. They showed that VCC and EC may achieve much lower latency than CC to reduce the task offloading time. However, this work does not address the related high deployment costs of Edge Computing when resources must be geographically dense.

Finally, in our previous work [14], we explored the feasibility of the substitution of EC by VCC when considering Wave as a communication technology. Even though Wave as a vehicle to infrastructure protocol has been largely adopted in literature, we rely in this work on 5G communications due to its high-performance potential in terms of coverage and achieved data rate and latency.

In conclusion, to the best of our knowledge, no previous study has clearly established the conditions under which VCC can fully replace EC in a 5G-enabled environment. This is despite extensive research into task offloading in 5G networks, such as the ones applying machine learning to EC task offloading or game theory for optimal strategies in large-scale EC scenarios [28]. Existing studies often view VCC as a complement to EC rather than as a standalone alternative.

Therefore, the objective of this work is not to add yet another strategy to this extensive collection. Instead, the aim is to conservatively analyze how VCC can take advantage of 5G wider coverage, better connectivity, and lower latency to effectively replace EC. We deliberately adopt simple strategies (more motivations on this choice can be found at the beginning of Section 4) to identify and investigate the parameters that govern the performance of an integrated three-tier Cloud architecture. The target is to analyze and understand the necessary conditions for complete EC replacement, especially in scenarios where the EC deployment may not be economically viable.

## 3. System architecture and networking

This section introduces the adopted architecture model and the 5G networking with the related parameters. The main objective is to define an architecture that can allow a fair comparison between EC and VCC. The fairness is guaranteed by adopting the same network settings for both computation paradigms.

### 3.1. System architecture

The system architecture considered in this study is illustrated in Fig. 1. It is based on a cellular access technology and includes six fundamental components.

Before introducing the architecture components, we first provide some definitions. The term **task** refers to the computation requested by an end-device user, which includes its execution on an external node and the feedback of the generated result (data or alternative code). A **task computation request** is the initial request sent by an end-user terminal, specifying the requirements for execution. **Task offloading** refers to the entire process of selecting a node, offloading the task, executing it, and returning the results. To maintain clarity, **task** can be used as a synonym for a request, while **task offloading** describes the complete execution process.

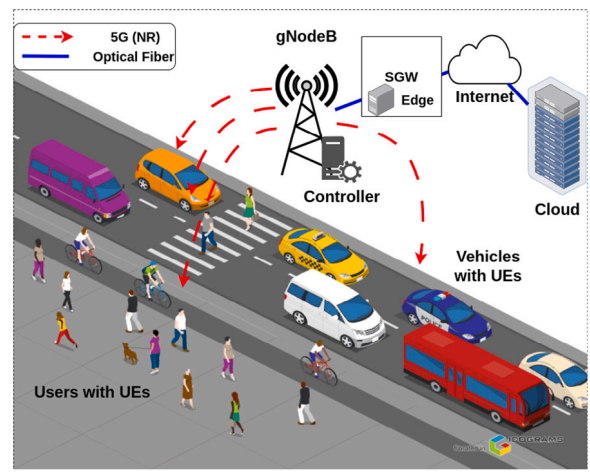

**Fig. 1.** Proposed communication architecture.

(1) **End-users** in need of offloading their tasks are identified as 5G User Equipment (UEs). UEs can be smartphones, smartwatches, laptops, tablets, etc. The UE of end-users is defined as Pedestrian-UE (PUE).
(2) A **gNodeB** (5G BS) with NR 5G technology (see 5G-LENA in Section 6) are macro BS, allowing a wide coverage area serving several moving vehicles. The large coverage allows a longer dwell time for vehicles, which ensures the completion of offloading requests, reducing failures.
(3) A **Controller**, implemented in the gNodeB, (i) keeps track of the vehicles available in the area (by collecting their beacons and recording their IP addresses) and (ii) decides the nodes where to offload incoming tasks. The detailed procedures executed in (i) and (ii) are given in Section 4.
(4) An **Edge server** is located in the Service Gateway (SGW). The SGW is a node at the Edge of the Core Network (CN), which ensures communication between the UEs and the external network parts. This location of the Edge server follows the ETSI standard for MEC architecture [29]. The Edge node has specific computational resources (such as CPUs and GPUs) and, if available, can compute offloaded tasks.
(5) **Vehicles** are equipped with on-board 5G UEs and computational resources capable of executing tasks [13] but assumed less powerful than the Edge's resources. The UE of vehicles is defined as Vehicle-UE (VUE).
(6) A **Cloud server** has a high amount of computational resources (they can be assumed to be infinite [30]), which outperforms those available at the Edge server.

The notion that Cloud resources are virtually unlimited is widely supported in the literature [31–33]. Thanks to the vast resource capacity of providers like Amazon or Google, users rarely experience task rejections. Offloading can be motivated by various factors, including energy savings and improvements in Quality of Experience (QoE) for resource-intensive applications. The QoE can be enhanced by reducing latency, which is achievable through Edge Computing (EC) and Virtual Cloud Computing (VCC).

### 3.2. Networking

This subsection describes the protocols and networking aspects of the architecture with the main functions and parameters.

#### 3.2.1. 5G infrastructure

We consider a 5G system architecture enabling 5G communication links between devices. The system is composed of a 5G cell (gNodeB or BS) employing an 8 × 8 tri-sector antenna. We use the 3GPP Release 15 for simulation purposes, such as in [34]. The channel is modeled at the range of FR1, i.e., at the sub-6 GHz frequency bands with a 5G-compliant channel frequency bandwidth of 200 MHz. Furthermore, the adopted numerology ($\mu = 2$) allows shorter latency. Lower numerologies provide better coverage but suffer from higher latency and reduced spectral efficiency. Higher numerologies enable lower latency and better spectral efficiency but have coverage limitations. The latter could reduce the Transmission Time Interval (TTI) too much, increasing packet loss.

Beam-forming is employed to increase the network capacity. The parameters above implement a macro BS that covers relatively distant vehicles in an urban scenario. The path-loss model is the urban macro line of sight (UMa LoS), suitable for urban scenarios. The UEs for end-users and vehicles are modeled similarly using a basic 1 × 1 array isotropic transmission antenna [35] with a transmission power (23 dBm for UEs) adapted to the case of study (task offloading on vehicles).

#### 3.2.2. Other networking aspects

The PUEs, as the VUEs, use the NR 5G lower layers integrating UDP/IP in the upper layers protocol stack. The UDP transport protocol is adopted as in [14] to make the communication lighter than TCP. Also, the **Edge server** stack is based on UDP/IP regarding upper layers, and the same above considerations for this choice are valid. The Cloud server is reachable over the Internet, implying higher network latency (35ms [6]) than those for reaching the Edge server. More details are reported in Table 2.

### 3.3. Beaconing for tracing vehicle availability

The Controller maintains a list of available vehicles to perform computation tasks. This list is generated based on the beacons sent by the vehicles, which announce their available computation resources. The Controller listens over appropriate multicast addresses (IPv6 or IPv4). Multicast addresses `beacon_address` and `request_to_offload_address` are used by vehicles and UEs to send their beacons and task offloading requests, respectively.

We assume two types of beacons: *periodic* and *aperiodic*. If a vehicle has computing resources available, it broadcasts a periodic beacon at a frequency of $f$. In our simulations, we set $f = 10$ Hz (i.e., one beacon every 100 ms). This frequency is consistent with the service requirements specified in 3GPP TS 22.186 [36], which recommends update intervals of 100 ms as a baseline for information exchange between vehicles and infrastructure in enhanced V2X scenarios. Upon receiving such a beacon, the Controller will add the corresponding vehicle to its list. If a vehicle is not available for computation of offloaded tasks (e.g., it is already processing a task), it does not send periodic beacons and hence does not appear in the Controller's list. In addition, when a controller sends a task to a specific vehicle, it removes that vehicle from its list. The vehicle will reappear in the list only after it transmits a new periodic or aperiodic beacon. Additionally, the Controller removes a vehicle from its list after a timeout (set to 500 ms in our simulations) if no new beacons are received. This timeout value is consistent with typical V2X beacon lifetimes and with the upper bounds for message relevance and freshness in vehicular safety applications [37].

Note that there might be a mismatch between the list of available vehicles in the Controller and the vehicles available in reality. This happens in the following cases:

- A vehicle with available resources just entered the cell at time $t$, but the next periodic beacon is sent at time $t + \epsilon$, where $\epsilon \in [0, 1/f]$. In this case, in the interval $[t, t+\epsilon]$, the resources on such a vehicle cannot be exploited.
- A vehicle available until time $t$, just exited the cell at time t. In this case, the Controller will remove such a vehicle from its list after the timeout, i.e., at instant $t + \epsilon'$, where $\epsilon' \in [0, \text{timeout}]$. During interval $[t, t + \epsilon']$, the Controller may still send tasks to that vehicle, however, without obtaining responses. Such tasks are thus doomed to fail (see red bar in Fig. 7).

## 4. Offloading strategies

We consider in this paper two basic offloading strategies, namely: *ECFirst* and *VCCFirst*. The motivations behind such choices rather than more advanced strategies are provided at the end of this section.

Let us assume that an end-user device (PUE) requires an external node to compute a task related to a local application. The PUE sends an offloading request to a Controller, which is co-located with the BS. The Controller is responsible for forwarding the task to the EC node or a VCC node. If the Controller finds that no computational resources are available in EC or VCC, CC is selected as a backup option, as it is assumed to have infinite available resources [30]. The node that computes the task sends the request to the dedicated queue if it exists, waiting for execution before returning the result to the sender.

It is worth noting that any issues related to privacy, energy consumption, and incentives for car owners to accept offloaded tasks in their vehicles are outside the scope of this paper and are addressed in the literature, such as [21,38].

Below, we provide definitions for the two offloading strategies:

- *ECFirst*: This scenario involves EC and CC for task offloading, and CC is considered a backup solution for managing task overflow. To reduce latency, the Controller prioritizes transferring tasks to the EC and resorts to sending tasks to CC only when the EC is operating at full capacity, i.e., upon saturation of the tasks queue of the Edge node. The queue in EC is handled with a First-In-First-Out (FIFO) policy.
- *VCCFirst*: This strategy includes only VCC and CC, suppressing the need for EC. Whenever the Controller receives a task to offload, it checks for the available vehicles in its internal list (maintained as explained in Section 3.3). Among the vehicles on the list with sufficient available resources, one is randomly selected for the task processing. Otherwise, if no vehicle is available for computation, the task is offloaded to the CC. From the Controller's point of view, vehicles are all equivalent since they are assumed to have the same resources and are at a reasonable distance from the gNB, ensuring coverage and feasibility of offloading.

If *VCCFirst* is adopted, vehicles send beacons to announce their presence and their available resource. To keep a conservative assessment, this paper considers that vehicles can handle only one task at a time. Therefore, a vehicle already serving a task offloading request rejects additional requests. This conservative assumption also guarantees that task offloading does not exhaust vehicle resources, whose main use should be driving safety and navigation.

In both ECFirst and VCCFirst strategies, CC is used as a backup solution. This means that the Controller sends to the Cloud all "overflow tasks", i.e., tasks that arrive when the queue in the EC is full (in the ECFirst strategy) or if no available vehicle with sufficient resources is found (in the VCCFirst strategy). According to the assumption that the Cloud has infinite capacity (Section 3.1), the Cloud always has available resources and will never reject a task.

Note that the strategies presented here are intentionally basic. Numerous optimization strategies have been proposed in the literature to perform task allocation in the VCC. The selection of vehicles could be done based on speed, vicinity to the edge node, or predicted trajectories instead of randomly selecting them. One could be tempted to adopt all such advanced strategies for the purpose of this paper. However, by doing so, findings will be tightly coupled to the particular strategy picked. This paper, instead, aims to provide more *conservative* findings by deliberately adopting basic strategies. Indeed, if VCC can replace EC already "in the worst case", i.e., with such basic strategies, one can confidently conclude that such replacement will also be possible with more advanced strategies, as they would obviously improve the performance of VCC. The performances shown in this paper are thus intended as pessimistic estimates of what one could achieve with their preferred state-of-the-art task allocation and vehicle selection algorithms.

It is worth noting that any strategy composed of VCC and EC is explicitly omitted. Indeed, using a Vehicular-Edge Computing strategy is not indicative since it behaves similarly to VCCFirst with EC instead of CC. However, to prove that the substitution of EC by VCC is feasible, the standalone paradigm (VCC or EC) with CC, used as a backup for managing the task overflow, is investigated.

## 5. Task and offloading time models

In this section, we present the task and offloading time models related to the different architecture components, namely CC, EC, and VCC.

### 5.1. Task model

A task $i$ is defined as a tuple $(W^i, S^i, R^i)$, where $W^i$ represents the workload (in Million Instructions [MI]), indicating the number of instructions required to execute task $i$. $S^i$ denotes the input task size (in [KB]), and $R^i$ represents the task processing result expressed in terms of the amount of data to be returned to the end user.

The offloading process unfolds as follows: once a given PUE decides to offload a task, it forwards the task with the corresponding tuple information to the BS. Then, the Controller within the BS selects the offloading destination (EC, VCC, or CC) based on the chosen strategy. In the selected destination node, the task is queued (only for EC) and executed, and the resulting output is fed back to the PUE.

### 5.2. Offloading time models

The offloading time models include the fundamental components of all the offloading procedures. The overhead related to task preparation is not considered part of the offloading procedure, as it occurs at the PUE device following the offloading decision procedure [39]. The preparation time, if present, could be neglected, as the transmission time is typically much larger. Moreover, as the security considerations are out of the scope of this paper, we neglected the time overhead inherent to the adopted security protocol. Finally, the protocol stack overhead is implicitly included in the model, as the NS3 simulations provide results including this overhead.

#### 5.2.1. Cloud offloading time

The main assumption here is that the CC disposes of infinite computational resources. Hence, every offloading request is instantly elaborated at the CC when received without generating any queuing time. $T_{CC}^i$ is the offloading time of task $i$ to the CC. It includes the time for the task computation request to be sent, processed, and sent back through the result packet to the PUE. It is composed of the following time components (in [seconds]):

- The first component is the *Uplink time*, noted as $T_{up, PUE-gNB}^i$, which is the time required for the task to reach the gNodeB through 5G NR technology from the PUE.

- The *Uplink Core Network (CN) time*, noted $T_{up, CN}^i$. It is time for the offloading request to pass from the gNodeB to the SGW (i.e., to reach the Internet). This time is assumed constant (i.e., fixed to 2 ms in the simulations [40]).
- Another component is the Internet latency, noted as $T_{up, Internet}^i$. It is the time delay that the offloading request takes to transfer across the Internet and reach the Cloud Server.
- After the overall *uplink time part*, there is the elaboration time of the request. The *Elaboration time* $T_{elab}^i$ is as follows: $T_{elab}^i = W^i/C_{CC}$, where $C_{CC}$ is the computational capacity of the Cloud Server, expressed in *Million Instructions Per Second* (MIPS). $W^i$ is the task workload measured in MI.
- Finally, the *Downlink time* is defined as the time required for the result to be sent back to the PUE who requested the task offloading. Similarly to the uplink time, it is composed by $T_{down, Internet}^i$ the latency from CC to the SGW, $T_{down, CN}^i$ the delay experienced from SGW to the gNB, and $T_{down, gNB-PUE}^i$ the transfer time from the gNodeB to the PUE who initiated for the offloading procedure.

Finally, the total task offloading time (a.k.a. task response time) of task $i$ to the CC, $T_{offloading, CC}^i$, can be expressed as follows:

$$T_{offloading, CC}^i = T_{up, PUE-gNB}^i + T_{up, CN}^i + T_{up, Internet}^i + T_{elab}^i + T_{down, Internet}^i + T_{down, CN}^i + T_{up, gNB-PUE}^i. \quad (1)$$

#### 5.2.2. Edge offloading time

As stated previously in Section 3.1, the Edge node is implemented at the SGW (see Fig. 1). The EC paradigm is composed of only one computational resource (one CPU or GPU), and its capacity, noted as $C_{EC}$, is measured in MIPS. If the computational resources are occupied, the EC employs a FIFO queue to manage offloading requests, ensuring fair task processing.

The task offloading time to the EC node, $T_{offloading, EC}^i$, is given as:

$$T_{offloading, EC}^i = T_{up, PUE-gNB}^i + T_{up, CN}^i + T_{queue}^i + T_{elab}^i + T_{down, CN}^i + T_{down, gNB-PUE}^i \quad (2)$$

where $T_{queue}^i$ is the time that task $i$ spends in the queue before the elaboration. $T_{elab}^i = \frac{W^i}{C_{EC}}$ is the elaboration time, and $W^i$ is the task workload measured in MI.

#### 5.2.3. Vehicular offloading time

Let $C_{VCC}$ be the computational capacity of a single vehicle in the VCC. For simplicity, in this work, all the vehicles are assumed to have the same computational resources since the objective is to avoid case-specific dependency bias. Moreover, we assume that the quantity of resources that the vehicle makes available for external tasks is considered as a portion of idle resources (i.e., not used for internal tasks such as in-vehicle driving tasks). Since the resources are assumed to be logically separated, external tasks do not impact internal tasks. In the case the resources are not homogeneous, the contribution of the elaboration time in the results would be proportional to this availability. Note that considering only a fraction of the actual vehicle resources is a conservative approach. In practice, processing time can be greatly reduced by leveraging more on-board power, as modern GPUs like the NVIDIA Tesla V100 reach 100 TeraFLOPS [41]. We will later show through simulations (in Section 6) that the impact of increasing computational resources is no more beneficial to the offloading time after a certain point. This happens since the communication time is, in this case, much longer than computation. The effect of vehicle resources heterogeneity will be investigated in our future works.

We assume that no queuing time is considered for vehicles since all vehicles can serve only one task at a time. The offloading time of a given task $i$ offloaded to a VUE is given by:

$$T_{offloading, VCC}^i = T_{up, PUE-to-gNB}^i + T_{up, gNB-VUE}^i +$$

$$T^i_{\text{elab}} + T^i_{\text{down, VUE-gNB}} + T^i_{\text{down, gNB-PUE}} \quad (3)$$

where:

- $T^i_{\text{up, PUE-gNB}}$ is the time required by the offloaded request to pass from the PUE end-device to the gNodeB. The opposite time for the *Downlink phase* is $T^i_{\text{down, gNB-PUE}}$.
- $T^i_{\text{up, gNB-VUE}}$ is the time required to send the request to a vehicle from the gNodeB. The opposite time for the *Downlink phase* is $T^i_{\text{down, VUE-gNB}}$.
- $T^i_{\text{elab}} = \frac{W^i}{C_{\text{VCC}}}$ is the elaboration time of task $i$, where $W^i$ is the task workload measured in MI.

## 6. Performance evaluation

This section shows that VCC can be a powerful alternative to EC for low-latency applications. The extensive simulations identify the various conditions crucial for this substitution, including user types, task workload, and VCC capacity in terms of the density of vehicles and onboard computation resources. The approach directs all task offloading requests through the gNodeB before reaching the vehicles using 5G links (Section 3).

### 6.1. Simulation and network environment

The Network Simulator version 3 (NS3) is adopted to implement the communication environment of the system shown in Section 3. The main simulation and network parameters are listed in Table 2. While being compliant with the 5G standard, several network parameters (such as bandwidth, transmission power, and operational frequency) are chosen empirically to ensure coherence with the requirements for latency and coverage. The simulation of the communication technology is provided by 5G-LENA [49], an NS3 module used for simulating the New Radio (NR) part of 5G (3GPP release 15 [34]). The module 5G-LENA[3] is supported by several research projects and a large and active community that ensures frequent updates and new releases.

The UEs of end-users and vehicles use the same physical channel with identical configuration parameters as the bandwidth, carrier, and transmission power. The urban pathloss model (Table 2) adopted is useful to create a large coverage. All the network model parameters are reported in Table 2. They follow the rationale presented in Section 3.2.1.

In simulations, for the sake of simplicity, limited-size tasks are considered, and the default task size is 4 KB inspired by Ref. [14] to enable comparison with the same. It is worth noting that a given task, even if it has a small size in KB, may be computation-intensive, i.e., it requires many instructions to be executed and, hence, a high workload. For instance, a simple "for loop" could have a size of less than 1 KB but could require a heavy workload in terms of Million Instructions [MI]. This justifies the choice of the proposed task size. In [13] the maximum tested is 1000 KB.

Vehicle mobility is simulated by the SUMO mobility simulator [50]. The mobility of the vehicles is based on the *Manhattan mobility model* [51], and all the obtained results are averaged over 9 simulation runs to ensure accurate values.

The considered scenario is the **total coverage scenario**. It is based on the Manhattan mobility model, implementing a single rectangular road composed of two lanes in the two possible directions of the road. The rectangle shape has length and width, respectively, on *x* and *y* axes of 600 m and 50 m, respectively. In this mobility scenario, the BS coverage is total. It is positioned at the center of the scenario with coordinates $(x, y, z) = (300, 25, 30)$m, where $z$ is the height of the BS (Table 2). This scenario allows vehicles to assume, on average, the maximum speed in the long line part of the road, making it useful for studying the effect of speed in experiments.

[3] 5G-Lena is supported by the *Centre Tecnològic de Telecomunicacions de Catalunya* (CTTC) a research institute in Spain focused on telecom and intelligent transportation tech.

**Table 2**
Simulation parameters.

| Parameter | Value |
|---|---|
| **Scenario** | |
| Cloud nodes | 1 |
| Number of Edge nodes | 1 (in the ECFirst scenario) or 0 (in the VCCFirst scenario) |
| Number of vehicles | 40 by default (up to 60) |
| Number of end-users | 8 [42] by default, (up to 100) |
| Simulation duration | 120 s |
| Cloud computation resources | $C_{CC}$ = 2356230 MIPS [43], $\infty$ processors |
| Edge computation resources | $C_{EC}$ = 749070 MIPS [44], 1 processor |
| VCC computation resources | $C_{VCC}$ = 71120 MIPS [45], 1 processor per vehicle |
| Task workload | $C_u$ = 500 MI [46] |
| Task size | $D_u$ = 4000 bytes [13] |
| Max queue length at EC | 100 packets (FIFO policy) |
| End-user offloading request rate | A request every 200ms [47] |
| Vehicle average speed | downtown traffic 13.1 km/h [48] |
| Used seeds for SUMO scenarios (total 9) | 0,1,2,3,4,6,7,8,9 |
| **Communication** | |
| gNodeB Height | 30 m [40] |
| UE Height | 1.5 m [40] (the average height of a device of a pedestrian) |
| Core Network latency | $T^i_{up,CN} = T^i_{down,CN} = 2$ milliseconds [40] |
| Internet latency | $T^i_{up,Internet} = T^i_{down,Internet} = 35$ milliseconds [6] |
| Operational frequency | 6 GHz (FR1 [34]) |
| Numerology | $\mu$ = 2 Section 6.1 |
| Tx Power of gNodeB | 20 dBm Section 6.1 |
| Tx Power of UE | 23 dBm Section 6.1 |
| Bandwidth | 200 MHz Section 6.1 |
| Antenna array gNB (three sectors) | 8 × 8 [35] |
| Antenna array UE (isotropic) | 1 × 1 [35] |
| Pathloss model | UMa LoS [35] |

Regarding pedestrians, they remain stationary during task offloading, which seems reasonable given its short period (0.5s). Even a "fast" pedestrian would likely move less than 1 meter within this time frame. The default processors selected are the *AMD Ryzen Threadripper 3990X (64 cores)* for the CC [43,52], *AMD Ryzen 9 3950X (16-core)* for the EC [44,52], and *ARM Cortex A73 (4-core)* for each vehicle [14,45]. These processor choices are made based on their market availability and computing power, as outlined in the existing literature on the computational capabilities of CC, EC, and VCC (embedded) [14,46].

The task workload is measured in Million Instructions [MI], and it is set at 500 MI. This assignment is related to *object recognition*. The practical implication of this choice is that a 500MI workload is representative of the variety of existing workloads. In fact, this value is classified as a "middle task", as reported in [46].

The acquired offloading times are compared to three reference application classes and their corresponding latency thresholds (Table 3). Those classes can cover most of the expected latency requirements in 5G networks. For instance, a user can be required to use augmented reality with object recognition and other related tasks. This is the case of extremely low and very low requirements. However, if an end-user needs some offloading of his antivirus, the requirements for task completion can be less strict. The impact of several parameters on performance is presented in the following section.

**Table 3**
Application classes and related latency requirements.

| Class name | Requirement | Example of applications |
| --- | --- | --- |
| Extremely Low Latency (LL$^{++}$) | $\leq 16$ ms | Augmented Reality [22] |
| Very Low Latency (LL$^{+}$) | $\leq 100$ ms | Augmented Reality [22] |
| Low Latency (LL) | $\leq 500$ ms | Antivirus [53] |

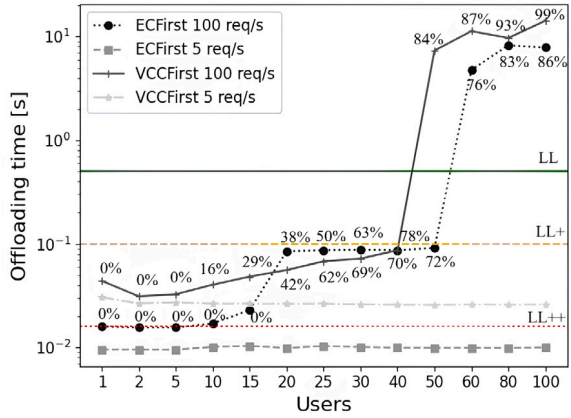

**Fig. 2.** Average offloading time ($T_{EC}^i, T_{VCC}^i$) of ECFirst and VCCFirst scenarios for different request rates. The reported percentage refers to the number of requests satisfied by Cloud Computing. This is reported for both strategies.

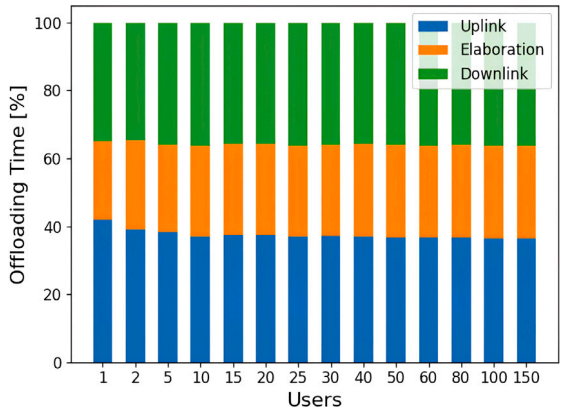

**Fig. 3.** Contribution of the different components of the offloading time in the VCCFirst scenario for the default request rate (percentage).

### 6.2. Impact of the number of end-users

In the figures, the three horizontal lines represent the latency requirements of the considered application classes (Table 3).

In Fig. 2, the offloading time (task completion time) is presented as a function of the number of users associated with the respective PUEs for both the adopted strategies, VCCFirst and ECFirst. The impact of the number of users is considered to be two different request rates, as reported in the legend of the figure. By increasing the number of users, the total request rate increases proportionally to them. The default request rate (Table 2) is related to the considered task of object recognition [46]. The default value of the request rate implies highly active users. The objective of this setting is to have continuous activity on the network to measure the performance of traffic offloading intense cases. Differently, if the offloading was considered non-continuous, it would not show the potential and capacity of the model in serving tasks. For the default task rate (5 requests/second), both strategies can manage 100 users and more, with no need to be helped by CC. This indicates that the resources and the network settings adopted (Table 2) are sufficient for this scenario, since 5G brings large network capacity. The figure shows, just for the curves related to the default values of the request rate, that the offloading time in EC is always lower than VCC. The results indicate that VCC can replace EC from 1 to 100 users for LL$^{+}$ and LL application requirements. *Extremely low latency*, LL$^{++}$, applications cannot be supported by VCC, but they can be satisfied only by EC. The figure also shows that even numerous users do not influence the offloading time because the network capacity is large. Comparing these results with the previous work in [14], for the same type of analysis, the degradation of performance is very visible. This is an advantage of 5G compared to Wi-Fi (IEEE 802.11p) for vehicular applications. Under these conditions, the contribution of the CC is negligible. The underlying phenomenon is the increase in channel utilization. The rationale behind this is that the number of offloaded tasks per second increases with the number of users. Hence, for 1 user, there are, on average, 5 requests/sec, and for 100 users, this number is 500 requests/sec. The objective here was to stress the network through higher request rates to show its limits. The 5G network loses in performance because of a higher request rate (an average of 10000 requests/sec for 100 users). This shows clearly that even 5G links could reach their capacity limitation in both strategies. For more clarity, an indication of the task overflow quantity is presented in the figure. The task overflow is managed by CC and a percentage of the requests that are sent to the CC is provided on the curve. The figure shows that for higher loads, the CC takes over since both the EC and VCC have too many requests to manage when alone.

Fig. 3 shows the contribution, in percentage, of the uplink, elaboration, and downlink time over the entire offloading time. The uplink and downlink parts offer similar contributions. The slightly minor contribution in terms of percentage over the offloading time is given by the elaboration time. This indicates that the number of adopted resources in the VCC is well-chosen after the parameters are fine-tuned.

This analysis concludes that the model exhibits resilience to high task loads and possesses the capability to manage tasks under high traffic conditions. This happens even for high numbers of users in only one gNB coverage area. Furthermore, both EC and VCC meet the established low latency requirements. While EC demonstrates superior performance for extremely low latency (approximately 10 ms), VCC displays potential for serving as a viable replacement for EC at around 30 ms. These results are supported by the ANOVA (Analysis of Variance) test presented in the following. ANOVA test is a statistical analysis. The analysis in this study is conducted to compare the impact of parameters between the two strategies, VCCFirst and ECFirst. The F-statistic in the Table 4 indicates the ratio between the variance resulting from the independent variable (users, workload) and the variance within the groups of offloading time observed values. This result indicates when a factor parameter has a significant effect. The P-mind denotes that the possibility of achieving the results observed is true, following at the same time the zero hypothesis. Low values suggest statistical importance. The class-sum considers the total variation in the data. The degree of freedom (DF), instead, indicates the number of independent comparisons that can be created, i.e., if the *x*-axis has $n = 10$ values, the possible comparisons are 9 ($n - 1$). In conclusion, the residual represents the part of the unexplained variance, which is not part of the variability responsible for the model. The ANOVA test shows the findings mentioned above in the case of user number variation. The correlation between the two strategy results is weak (F = 0.0045, p = 1.0) for the default request rate (5 requests/s) for strategy results. However, if the system is overloaded (100 requests/s), the sensitivity of the offloading time to the number of users becomes higher (F = 6.311177, p = 0.001174) due to the high channel utilization.

**Table 4**
ANOVA results for users (request rate is firstly 5 requests/second and after 100 requests/second) and workload and relative significance.

| | Parameter | Comment | sum_sq | df | F | PR(>F) |
|---|---|---|---|---|---|---|
| C(Users) | Users(low rate) | No significant effect | 0.000008 | 12.0 | 0.0045 | 1.0 |
| Residual | – | – | 0.001814 | 13.0 | – | – |
| C(Users) | Users(high rate) | Significant effect | 401.167688 | 12.0 | 6.311177 | 0.001174 |
| Residual | – | – | 68.861688 | 13.0 | – | – |
| C(Workload) | Workload | No significant effect | 0.016559 | 12.0 | 0.799279 | 0.647773 |
| Residual | – | – | 0.022444 | 13.0 | – | – |

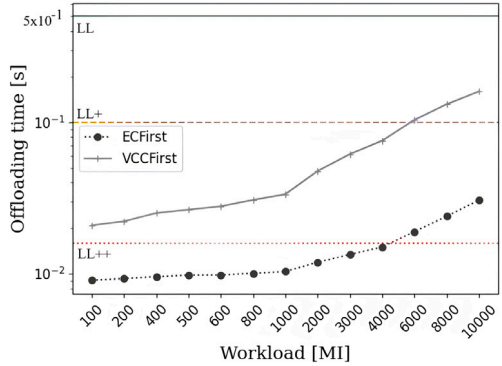 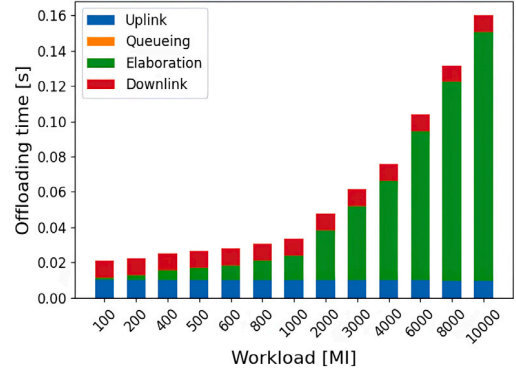

(a) ECFirst and VCCFirst offloading time.

(b) The VCC offloading time for the *VCCFirst* strategy.

**Fig. 4.** Comparison of offloading times on workload variation.

*6.3. Impact of task workload*

This subsection presents the effect of the workload on the offloading time. An important impact on the offloading time, in particular, the computation time, is expected.

Fig. 4(a) shows the offloading time as a function of the task workload for each strategy.

According to [46], the workload can be divided into three categories according to the task complexity, starting from 100 MI to 9784 MI. The task workload is 500 MI in the results in Fig. 2 and Fig. 6. The application related to this workload is the *object recognition* [46]. Fig. 4(a) shows the offloading time for the two strategies. The results strongly suggest that the EC can handle all the possible task workloads and stay within the deadline of 16 ms until 4000 MI. It can serve the remainder of the requirements after this value. This is not the case with *VCCFirst*.

The VCC meets the $LL^+$ and LL requirements from 100 MI to 6000 MI. After 6000 MI, the VCC can fulfill the LL requirements, while the EC can still satisfy $LL^+$ and LL. This happens without relying on CC. Fig. 4(b) presents the different components of the offloading time. As expected, the elaboration time for the VCC increases proportionally to the workload. The conclusion from these results is that for all the presented task workloads, low latency is satisfied by both VCC and EC, which means that VCC can replace EC even for this analysis. The conclusion here is that for all the presented task workloads, the low latency is satisfied by both VCC and EC, which means that VCC can replace EC even for this analysis. The ANOVA test presented in Table 4 shows that the tested workload range has no significant effect on the performance of both strategies.

*6.4. Impact on the density of vehicles*

In this subsection, results are presented only for VCCFirst. EC analysis would not add any meaningful information. In Fig. 5, the bottom *x*-axis represents all the simulated vehicles in the scenario, all evolving under the coverage of the BS. This means that on the bottom *x*-axis all the vehicles compose the VCC since they are all in BS's coverage area. The top *x*-axis shows the mean number of vehicles in the VCC that is used for task offloading, averaged over the simulation time and the several simulation runs (Table 2). This set corresponds to the vehicles from which the Controller has received fresh beacons, saying that the vehicular resources are available and exploited. This number of used vehicles in VCC is calculated as the average over all the simulation offloaded tasks of the number of vehicles involved in the task offloading. As expected, this average increases with the number of vehicles. At the same time, the usage of CC, expressed by the percentage in the graph, diminishes. These results indicate that with enough resources in the VCC, the last improves the response time of tasks, and it allows better performance regarding cases of great usage of CC. The CC use begins to be negligible from 10 vehicles in the scenario, and it decreases more and more by increasing the amount of the capacity of the VCC.

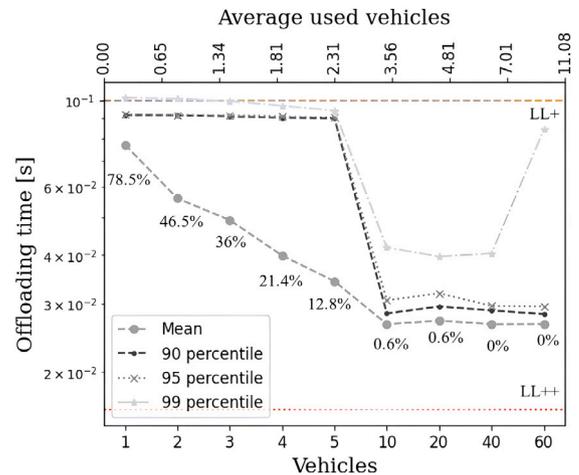

**Fig. 5.** Offloading time of VCCFirst scenario. Impact of the number of vehicles used vehicles, and vehicles in VCC. The reported percentage refers to the number of requests satisfied by Cloud Computing. Several curves indicate the 90th, 95th, and 99th percentiles of all the offloading values over the simulation.

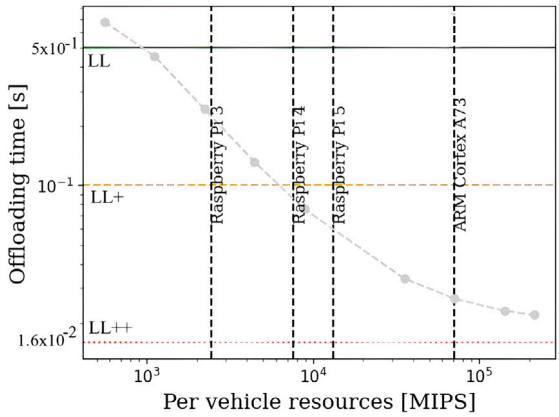

**Fig. 6.** Offloading time of VCCFirst scenario. Impact of computational capacity $C_{VCC}$ of each vehicle.

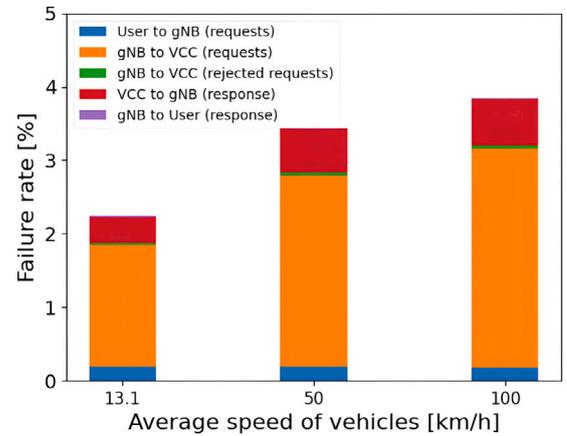

**Fig. 7.** Failure rate in percentage of the total offloading requests per diverse speeds.

In the same figure, the curves correspond to the 90th, 95th, and 99th percentiles. These curves are farther from the average offloading time values. In fact, in CC scenarios, when the variability of the offloading times is large, higher values correspond to these curves. As soon as the usage of VCC increases, these curves get closer to the average, showing reduced variability.

Another significant observation to note is that the 99th percentile is pretty high compared to the average, something one would expect given that CC is used so infrequently, only very rarely, that there are outliers in offloading times. The percentiles presented here help highlight the extreme situations and offer a glimpse into the range of system performance and stability. This allows for the prevention of worst-case scenarios.

This analysis shows that independently of the vehicle number within the tested range, the VCCFirst strategy is sufficient to ensure the $LL^+$ latency. This is because even for low vehicle density, the Cloud can satisfy the overflow of requests under 100 ms. However, when the number of vehicles is sufficient, the offloading time stabilizes at lower values, ensuring a better QoS. The latter can be explained by the proximity of vehicles to end-users, compared to CC, leading to shorter communication time.

### 6.5. Impact of the computational resources deployed into vehicles

This subsection presents how the variation of resources on vehicles impacts the offloading time.

In Fig. 6, a CPU with capacity $C_{VCC} = 71120$ MIPS is considered, which is the *ARM Cortex A73*, and this is compared to Raspberry Pi 3, Raspberry Pi 4, Raspberry Pi 5 CPUs [54,55]. The *x*-axis values are $C_{VCC} \cdot (1/128, 1/64, 1/32, 1/16, 1/8, 1/2, 1, 2, 3)$. The analysis of the increase in the resources reveals that at a certain point, more resources are very slightly beneficial to the performance because the quality ratio, price, and number of resources are no longer convenient. The figures show that by increasing the vehicle resources further (beyond the ARM Cortex A73 capacity), the benefits of offloading time would not be significant. Even though elaboration time tends to be zero, communication time (uplink and downlink) has a larger impact on the overall offloading time. The CC is never used, which means that the adopted CPU is large enough to satisfy the offloading. This analysis concludes that the adopted computational power makes VCC capable of satisfying $LL^+$ and LL and very close to $LL^{++}$ requirements. The latter strengthens the advantage of VCC as a replacement for EC.

### 6.6. Impact of the speed

In this part, a scenario with partial coverage of the gNB is adopted. The *x*-axis length of the SUMO scenario is twice the length of the rectangle of the **total coverage scenario**. The extended rectangular simulation area (1200 m *x* 50 m) provides sufficient road length for vehicles to accelerate and reach higher speeds, allowing us to test performance under various mobility conditions.

Fig. 7 shows the failure rate for the studied speeds using as default parameters the ones in Table 2.

The adopted values of speed are $\{13.1, 50, 100\}$km/h. The figure distinguishes the percentage of failure calculated regarding the total number of requests for offloading in the simulation.

The first observation is that the failure rate increases with speed. The first part of communication (user to gNB) is constant because the requests in the network do not change numbers. The second part of the failure is related to the failed requests related to the communication from the gNB to the VCC. This means that in this part of the communication, mobility (speed in this case) is crucial for success in the offload. After that, there is very little rejection failure in vehicles, which means that vehicles are occupied when they receive a request. This can happen because the list in the Controller, which contains available vehicles, is not exactly the real situation. This is due to the difference between the instant of the last update and the real use of the resources. Observe that increasing the signaling of the vehicles' beacons to reduce rejection failure is not needed. Firstly, because this value is negligible, and secondly, because it corresponds to a realistic phenomenon. The last part of the failure (VCC to gNB) is due to the answer after taking an offloading request in charge and answering to the gNB. The failure increases slightly, even if the speed increases considerably. This means that the failure related to an offloading request received by a vehicle when the vehicle goes out of coverage is very low. The last part of failure (gNB to the user) is negligible, and it is related to the normal function of the network, i.e., it depends on the error model of the 5G network. Another reason is that the distribution (the load of the network) is less in the response part of the task offloading. This is because of the offloading request lost in the previous steps of the communication, and the elaboration of tasks in one node is a part that makes the output slower than the input.

The underlying effect is due to several factors. Firstly, vehicle velocity impacts the channel quality in both uplink and downlink, resulting in the degradation of communication quality. This leads to higher failure rates, particularly at the gNB coverage border, which is emphasized by interference and noise phenomena. Furthermore, those effects create inconsistencies between the real number of vehicles under coverage available for offloading and the vehicles in the Controller list (see Section 3). The inconsistency occurs when vehicles close to the cell border signal their availability while immediately leaving the gNB coverage, affecting the number of failures in both uplink and downlink between gNB and vehicles. This phenomenon is magnified by the vehicles' velocity.

## 7. Cost discussion

This section presents an evaluation of the economic interest of a Network Operator (NO) relying on VCC as an alternative to EC. A deep and technical cost assessment would be out of the scope of this work and would deserve a separate paper in a separate journal focused on business management. The assessment of this section will thus purposely remain high-level. For simplicity, the evaluation is performed on a generic cell.

### 7.1. CAPEX

The part of the cost related to the equipment is the Capital Expenditure (CAPEX). The NO bears the CAPEX for the EC, but not for the VCC. Indeed, the NO must buy computational resources and install them at the Edge in the case of EC. In the case of VCC, instead, the NO opportunistically uses the resources already deployed in the vehicles.

Let $L_{\text{EC-CPU}}$ be the lifespan of a computational resource (CPU, GPU, …) deployed at the Edge node, i.e., the time during which the node is supposed to work before a significant performance degradation and the need for replacement. Let $Y$ denote the investment duration, expressed in *years*, along which the adoption of VCC versus EC is evaluated. The CAPEX incurred by the NO to deploy computational resources at the Edge is:

$$c_{\text{CAPEX-EC}} \geq c_{\text{EC-CPU}} \cdot \left\lceil \frac{Y}{L_{\text{EC-CPU}}} \right\rceil \qquad (4)$$

The term $c_{\text{EC-CPU}}$ is the market price for a single computational resource as CPU. The factor $\lceil Y/L_{\text{EC-CPU}} \rceil$ indicates how many times the NO is expected to renew the computational capacity in EC over the investment duration, $Y$. The sign $\geq$ denotes the fact that the cost of deployment is higher than just considering the CPU, as other components, such as RAM, disk, etc., are also needed.

### 7.2. OPEX

Let us now present the Operational Expenditure (OPEX) costs. The NO incurs costs $c_{\text{EC-req}}$ for each task offloaded using the EC. This accounts for the cost of the energy spent in EC for the entire offloading procedure, i.e., communication and computation required by the task. About the VCC, the NO incurs cost $c_{\text{VCC-req}}$ per offloading request. Indeed, a car owner (or the car manufacturer, depending on the business model that will develop around VCC) accepts to make the spare computational capacity available in the car only against payment. Such a payment is performed by the NO for each processed task. Moreover, the NO spends some energy transmitting the task to the car and receiving the response over the air, which also has a cost. The value $c_{\text{VCC-req}}$ is the cost borne by the NO for both payment and energy to send the task to the VCC and receive back the result from a VCC node.

In the case of EC, the OPEX also includes maintenance, which consists of hardware and software repairs to keep the EC functioning correctly. The NO does not instead incur any maintenance cost in the case of VCC, as it does not own any resources. Therefore, the OPEX for EC is:

$$c_{\text{OPEX-EC}} = c_{\text{EC-req}} \cdot R \cdot U \cdot Y \cdot \alpha + c_{\text{EC-main}} \cdot Y \qquad (5)$$

where $R$ denotes the average task offloading requests of a user per second. The term $U$ is the average number of users present in a generic cell. $\alpha$ is the number of seconds users actively send offloading tasks in one year. It is assumed that users have 15 h of activity per day. $c_{\text{EC-main}}$ is the annual cost of maintenance of an EC node.

Instead, the OPEX born by the NO related to the VCC is just how much the NO pays to send the offloading requests to car owners (or manufacturers:)

$$c_{\text{OPEX-VCC}} = c_{\text{VCC-req}} \cdot R \cdot U \cdot Y \cdot \alpha. \qquad (6)$$

$$\begin{aligned} c_{\text{EC}} - c_{\text{VCC}} &= c_{\text{CAPEX-EC}} + c_{\text{OPEX-EC}} - (0 + c_{\text{OPEX-VCC}}) \\ &= c_{\text{CAPEX-EC}} + c_{\text{OPEX-EC}} - c_{\text{OPEX-VCC}} \\ &\underset{\text{Eqn. (4), (5), (6)}}{\geq} c_{\text{EC-CPU}} \cdot \left\lceil \frac{Y}{L_{\text{EC-CPU}}} \right\rceil + c_{\text{EC-main}} \cdot Y \\ &\quad + (c_{\text{EC-req}} - c_{\text{VCC-req}}) \cdot R \cdot U \cdot Y \cdot \alpha \end{aligned} \qquad (7)$$

where $c_{\text{EC}}$ and $c_{\text{VCC}}$ are the total costs of EC and VCC, respectively, over the investment period, which include both CAPEX and OPEX (as explained before, the CAPEX for VCC is 0).

These savings are positive if

$$c_{\text{VCC-req}} \geq c_{\text{EC-req}} + \beta \qquad (8)$$

where

$$\beta = \frac{c_{\text{EC-CPU}} \cdot \lceil Y/L_{\text{EC-CPU}} \rceil / Y + c_{\text{EC-main}}}{R \cdot U \cdot \alpha} \qquad (9)$$

In other words, VCC remains economically more convenient than EC, even in cases where the NO pays the car owner (or manufacturer) a price per request that is larger than the cost per request that it would bear in EC. The term $\beta$, is called *VCC bonus*, indicates how much larger the payment can be for the VCC to remain competitive with EC. As expected, this bonus is larger when the cost of installing $c_{\text{EC-CPU}}$ and maintaining $c_{\text{EC-main}}$ resources in EC is larger. This bonus is also high when request rate $R$, users $U$ offloading, and activity period $\alpha$ are low. This suggests that VCC can be a suitable alternative to EC in an initial phase when the penetration rate of low-latency offloading applications is low. Before installing EC, the NO could satisfy the requirements of such applications via VCC. EC would then be installed only after a critical mass of users started to require low-latency offloading capabilities.

### 7.3. Cost numerical results

In this subsection, an assessment of the cost is presented. The adopted parameters are reported in Table 5, and their values are representative for conducting the cost analysis. The chosen capacity cost for vehicular CPU and Edge CPU is based on market prices available at the time of the study. These values evolve over time as the market does since the prices are subject to variation over time. The proposed analysis reflects the perspective of the main stakeholder and investor, i.e., the Network Operator (NO). This stakeholder owns the internet and communication infrastructures. However, a dedicated and structured framework is required to better analyze the relevant parameters for other stakeholders. Such a model must focus entirely on economic analysis, modeling the benefits and decisions for all stakeholders. This aspect is beyond the scope of this paper. Instead, in future work, we propose a game-theoretical framework that considers complementary parameters, such as network delay, task deadline, and energy consumption related to the entire offloading process. Such a model would provide a better suitable framework for understanding how investment decisions could be beneficial. The numerical results presented in this study reflect the perspective of the main investor in Edge Computing (EC), the NO.

Fig. 8 shows the total cost of EC and VCC. All the plots represent the total cost of the EC (left bars), with the total cost of the VCC (right bars). On the *x*-axis, three values of $\beta$ are provided. According to the Eq. (8), the savings obtained by the use of VCC and CC are positive if the $\beta$ is positive or null. This indicates that VCC is more profitable. In these

**Table 5**
Adopted values for the cost analysis.

| Variable | Meaning | Value |
|---|---|---|
| $c_{EC}$ | Total expenditure in EC | Eq. (7) |
| $c_{VCC}$ | Total expenditure in VCC | Eq. (7) |
| $c_{CAPEX\text{-}EC}$ | Expenditure of the EC computational equipment | Eq. (4) |
| $c_{CAPEX\text{-}VCC}$ | Expenditure of the VCC computational equipment | 0 (7) |
| $c_{OPEX\text{-}EC}$ | Operational expenditure in EC | Eq. (5) |
| $c_{OPEX\text{-}VCC}$ | Operational expenditure in VCC | Eq. (6) |
| $c_{EC\text{-}CPU}$ | Cost per one EC CPU | 700[$] [56] |
| $L_{EC\text{-}CPU}$ | EC CPU lifespan | 3 years [57] |
| $c_{EC\text{-}main}$ | Annual maintenance cost per EC node, assuming one technician maintains 50 nodes per year (cost is 1/50 of technician's salary) | 1368.46[$] = (68423/50)[$] [58] |
| $c_{EC\text{-}req}$ | Offloading cost per request in EC | $2 \cdot 10^{-5}$ $/request [59] |
| $c_{VCC\text{-}req}$ | Offloading cost per request in VCC | Same as in EC |
| $Y$ | Investment duration (operational period for EC and VCC) | [1, 3, 5] years |
| $\alpha$ | The number of seconds users actively send offloading tasks in one year | Active 15 h per day |
| $R$ | Request rate per user | 5 requests/s (Table 2) |
| $U$ | Number of users in one day in one 5G cell who offload their tasks | 100 (Table 2) |
| $\beta$ | VCC bonus, indicates how much larger the payment can be for the VCC to remain competitive with EC | Eq. (9)[$/request] |

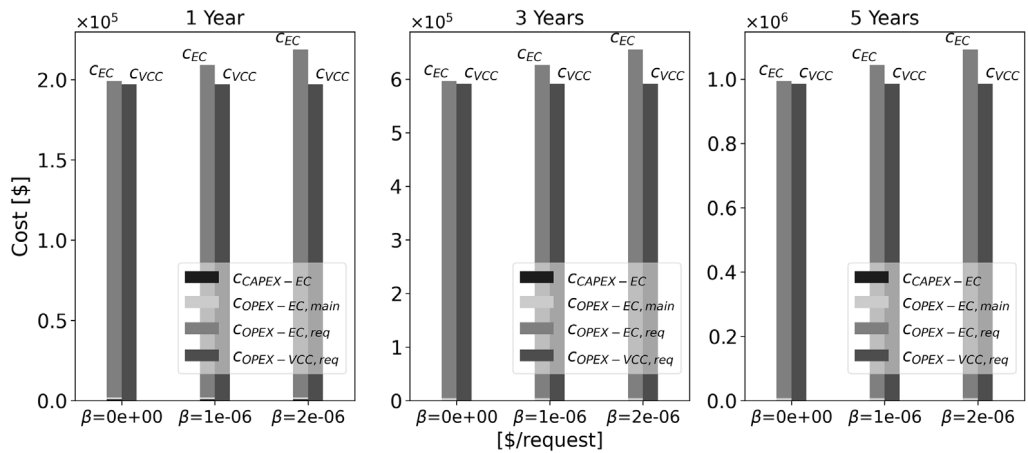

**Fig. 8.** Total cost of the EC compared with VCC for 1, 3 and 5 operational years.

figures, the conservative case ($\beta = 0$) is represented. In fact, the cost per request of VCC is $c_{VCC} = c_{EC\text{-}req}$ as reported in Table 5. Then, Eq. (8) becomes: $\beta \geq 0$. The findings support the theory in Eq. (8). In fact, for at least $\beta = 0$ every year, the savings are positive. Indeed, the gap between EC and VCC is precisely the sum of the maintenance cost plus the Capital Expenditure (CAPEX). Hence, the findings confirm that the real barrier of the EC deployment is related to the initial costs, which is evident even in the worst conditions, i.e., $\beta = 0$ and $c_{VCC} = c_{EC\text{-}req}$.

Furthermore, in Table 6, the influence of the $c_{CAPEX\text{-}EC}$ and $c_{EC\text{-}main}$ result weak, while the OPEX related to requests for CC and VCC results in almost all the expenditure. Even though the findings show that deployment and maintenance are limited to the given parameters, if the NO has to invest in hundreds or thousands of nodes, the influence of the cost of deployment would be non-negligible. The percentage of EC would remain the same. However, Fig. 9 and Table 7 show that using VCC is the best choice even in *unfavorable scenarios*, such as poorly planned EC investments or changing user needs leading to a market shift. The market substitution will be brought about by technological innovation, cheaper alternatives, or changes in the user's interests. The findings of this analysis confirm that just the use of VCC is a safer choice with respect to EC, in the case the offloading requests are the 1% of the ones in Fig. 9 and Table 5. This study concludes that the VCC is a safer investment option and, at the same time, cheaper than the EC.

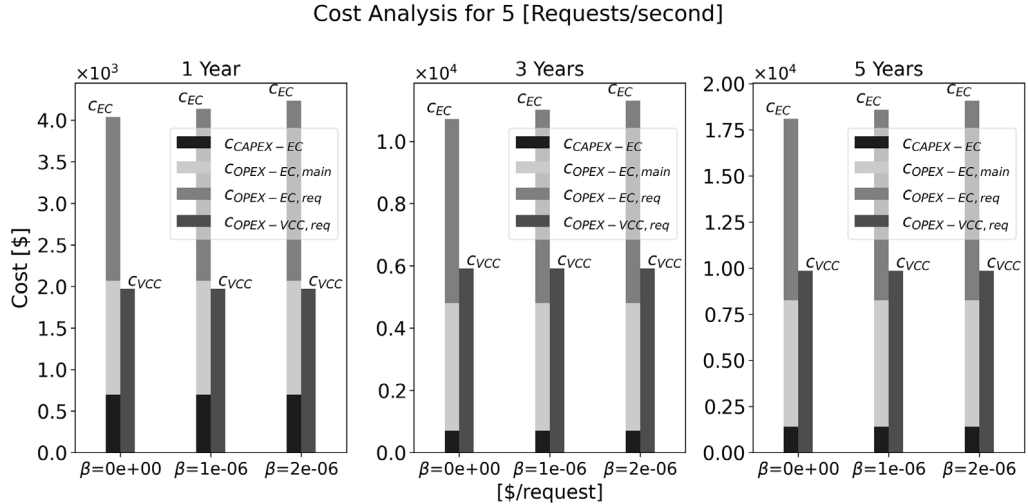

**Fig. 9.** Total cost of the EC compared with VCC for 1, 3, and 5 operational years for 1% of the total requests rate in the previous cost figure, i.e., 500 requests/second before and 5 requests/second here.

**Table 6**
Distribution of costs for EC and VCC in the first year, expressed in percentage.

| $\beta$ | $c_{\text{CAPEX-EC}}$ (%) | $c_{\text{EC-main}}$ (%) | $c_{\text{OPEX-EC,req}}$ (%) | $c_{\text{OPEX-VCC,req}}$ (%) |
|---|---|---|---|---|
| 0e+00 | 0.35 | 0.69 | 98.96 | 100.00 |
| 1e−06 | 0.33 | 0.65 | 99.01 | 100.00 |
| 2e−06 | 0.32 | 0.63 | 99.05 | 100.00 |

**Table 7**
Distribution of costs for EC and VCC in the first year, expressed in percentage for the 1% of offloading requests.

| $\beta$ | $c_{\text{CAPEX-EC}}$ (%) | $c_{\text{EC-main}}$ (%) | $c_{\text{OPEX-EC,req}}$ (%) | $c_{\text{OPEX-VCC,req}}$ (%) |
|---|---|---|---|---|
| 0e+00 | 17.33 | 33.88 | 48.79 | 100.00 |
| 1e−06 | 16.92 | 33.07 | 50.01 | 100.00 |
| 2e−06 | 16.52 | 32.30 | 51.18 | 100.00 |

## 8. Conclusion

This study illustrates that Vehicular Cloud Computing (VCC) is a cost-effective replacement for Edge Computing (EC), and can satisfy the requirement of low-latency applications. Indeed, failure rates remain very low (< 4%), even in high vehicle mobility scenarios. Moreover, even relatively few utilized vehicles, e.g., more than 4, can already match the needs of task offloading, except with very high request rates, where Cloud Computing must be used as a backup. A cost analysis shows that VCC can bring considerable savings to network operators when deploying EC. However, EC remains irreplaceable in extreme cases when latency requirements fall below 16 ms.

In the analysis, simple conservative task allocation strategies are employed. We chose to do so since we aimed to assess the benefits of VCC in a conservative setting. Future work will focus on devising strategies that intelligently allocate tasks based on their individual deadlines and factor in energy consumption.

## CRediT authorship contribution statement

**Rosario Patanè:** Writing – review & editing, Writing – original draft, Visualization, Validation, Supervision, Software, Resources, Project administration, Methodology, Investigation, Funding acquisition, Formal analysis, Data curation, Conceptualization. **Nadjib Achir:** Writing – review & editing, Writing – original draft, Validation, Software, Methodology, Investigation, Formal analysis, Data curation, Conceptualization. **Andrea Araldo:** Writing – review & editing, Writing – original draft, Visualization, Validation, Supervision, Methodology, Investigation, Formal analysis, Data curation, Conceptualization. **Lila Boukhatem:** Writing – review & editing, Writing – original draft, Visualization, Validation, Supervision, Methodology, Formal analysis, Conceptualization.

## Declaration of competing interest

The authors declare the following financial interests/personal relationships which may be considered as potential competing interests: Rosario Patane' reports was provided by Paris-Saclay University. PhD Student reports a relationship with Paris-Saclay University that includes:. Previous student at University of Catania, no conflicts known to declare If there are other authors, they declare that they have no known competing financial interests or personal relationships that could have appeared to influence the work reported in this paper.

## Acknowledgments

This work is supported by Labex DigiCosme, France.

## Data availability

The link to data is provided into the paper, they can be found on github repository.

**References**

[1] Boonyarith Saovapakhiran, Michael Devetsikiotis, Enhancing computing power by exploiting underutilized resources in the community cloud, in: 2011 IEEE International Conference on Communications, ICC, 2011, pp. 1–6, http://dx.doi.org/10.1109/icc.2011.5962544.

[2] Sohan Singh Yadav, Zeng Wen Hua, Cloud: A computing infrastructure on demand, in: 2010 2nd International Conference on Computer Engineering and Technology, vol. 1, 2010, pp. V1–423–V1–426, http://dx.doi.org/10.1109/ICCET.2010.5486068.

[3] AMAZON, The AWS blog: The first five years, 2009, https://aws.amazon.com/blogs/aws/aws-blog-the-first-five-years/, Retrieved 7 2025.

[4] Google Cloud Platform, Your next home in the cloud, 2025, https://cloud.google.com/blog/products/gcp/google-cloud-platform-your-next-home-in-the-cloud/?hl=en, Retrieved 7 2025.

[5] Zongkai Liu, Penglin Dai, Huanlai Xing, Zhaofei Yu, Wei Zhang, A distributed algorithm for task offloading in vehicular networks with hybrid fog/cloud computing, IEEE Trans. Syst. Man Cybern.: Syst. 52 (7) (2022) 4388–4401, http://dx.doi.org/10.1109/TSMC.2021.3097005.

[6] Verizon, IP latency statistics, 2025, https://www.verizon.com/business/terms/latency/, Retrieved 7 2025.

[7] Arif Ahmed, Ejaz Ahmed, A survey on mobile edge computing, in: 2016 10th International Conference on Intelligent Systems and Control, ISCO, 2016, pp. 1–8, http://dx.doi.org/10.1109/ISCO.2016.7727082.

[8] Li Lin, Xiaofei Liao, Hai Jin, Peng Li, Computation offloading toward edge computing, Proc. IEEE 107 (8) (2019) 1584–1607, http://dx.doi.org/10.1109/JPROC.2019.2922285.

[9] Rosario Patanè, Andrea Araldo, Tijani Chahed, Diego Kiedanski, Daniel Kofman, Coalitional game-theoretical approach to coinvestment with application to edge computing, in: 2023 IEEE 20th Consumer Communications & Networking Conference, CCNC, 2023, pp. 517–522, http://dx.doi.org/10.1109/CCNC51644.2023.10060093.

[10] Mario Gerla, Vehicular cloud computing, in: 2012 the 11th Annual Mediterranean Ad Hoc Networking Workshop, Med-Hoc-Net, 2012, pp. 152–155, http://dx.doi.org/10.1109/MedHocNet.2012.6257116.

[11] Iyshwarya Ratthi Kannan, Yogameena Balasubramanian, Saravana Perumaal Subramanian, Menaka Kandhasamy, Sivasankari Ramesh, CDANet: Computer vision based automatic car damage analysis, in: Proceedings of the Fourteenth Indian Conference on Computer Vision, Graphics and Image Processing, ICVGIP '23, New York, NY, USA, 2024, Association for Computing Machinery, ISBN: 9798400716256, http://dx.doi.org/10.1145/3627631.3627662.

[12] NVIDIA Danny Shapiro, Tesla unveils top AV training supercomputer powered by NVIDIA A100 GPUs, 2021, https://blogs.nvidia.com/blog/2021/06/22/tesla-av-training-supercomputer-nvidia-a100-gpus/, Retrieved 7 2025.

[13] Hongli Zhang, Qiang Zhang, Xiaojiang Du, Toward vehicle-assisted cloud computing for smartphones, IEEE Trans. Veh. Technol. 64 (12) (2015) 5610–5618, http://dx.doi.org/10.1109/TVT.2015.2480004.

[14] Rosario Patanè, Nadjib Achir, Andrea Araldo, Lila Boukhatem, Can vehicular cloud replace edge computing? in: WCNC 2024 - IEEE Wireless Communications and Networking Conference, Dubai, United Arab Emirates, 2024.

[15] FCC takes spectrum from auto industry in plan to "supersize" WiFi, 2020, https://arstechnica.com/tech-policy/2020/11/fcc-adds-45mhz-to-wi-fi-promising-supersize-networks-on-5ghz-band/, Retrieved 22 2024.

[16] Mohammad Aazam, Saif ul Islam, Salman Tariq Lone, Assad Abbas, Cloud of things (CoT): Cloud-fog-IoT task offloading for sustainable internet of things, IEEE Trans. Sustain. Comput. 7 (1) (2022) 87–98, http://dx.doi.org/10.1109/TSUSC.2020.3028615.

[17] Jaber Almutairi, Mohammad Aldossary, A novel approach for IoT tasks offloading in edge-cloud environments, J. Cloud Comput. 10 (1) (2021) 28, http://dx.doi.org/10.1186/s13677-021-00243-9.

[18] Marco Malinverno, Josep Mangues-Bafalluy, Claudio Ettore Casetti, Carla Fabiana Chiasserini, Manuel Requena-Esteso, Jorge Baranda, An edge-based framework for enhanced road safety of connected cars, IEEE Access 8 (2020) 58018–58031, http://dx.doi.org/10.1109/ACCESS.2020.2980902.

[19] Xiang Ju, Shengchao Su, Chaojie Xu, Haoxuan Wang, Computation offloading and tasks scheduling for the Internet of Vehicles in edge computing: A deep reinforcement learning-based pointer network approach, Comput. Netw. (ISSN: 1389-1286) 223 (2023) 109572, http://dx.doi.org/10.1016/j.comnet.2023.109572.

[20] Yuxuan Sun, Xueying Guo, Jinhui Song, Sheng Zhou, Zhiyuan Jiang, Xin Liu, Zhisheng Niu, Adaptive learning-based task offloading for vehicular edge computing systems, IEEE Trans. Veh. Technol. 68 (4) (2019) 3061–3074, http://dx.doi.org/10.1109/TVT.2019.2895593.

[21] Fangming Liu, Jian Chen, Qixia Zhang, Bo Li, Online MEC offloading for V2V networks, IEEE Trans. Mob. Comput. 22 (10) (2023) 6097–6109, http://dx.doi.org/10.1109/TMC.2022.3186893.

[22] Ayoub Ben Ameur, Andrea Araldo, Francesco Bronzino, On the deployability of augmented reality using embedded edge devices, in: 2021 IEEE 18th Annual Consumer Communications & Networking Conference, CCNC, 2021, pp. 1–6, http://dx.doi.org/10.1109/CCNC49032.2021.9369590.

[23] R.A. Isaac, P. Sundaravadivel, V.S.N. Marx, et al., Enhanced novelty approaches for resource allocation model for multi-cloud environment in vehicular Ad-Hoc networks, Sci Rep 15 (2025) 9472, http://dx.doi.org/10.1038/s41598-025-93365-y.

[24] Hengwei Liu, Ni Tian, Deng-Ao Song, Long Zhang, Digital twin-enabled multi-service task offloading in vehicular edge computing using soft actor-critic, Electronics 14 (4) (2025).

[25] Chen Chen, Yini Zeng, Huan Li, Yangyang Liu, Shaohua Wan, A multihop task offloading decision model in MEC-enabled internet of vehicles, IEEE Internet Things J. 10 (4) (2023) 3215–3230, http://dx.doi.org/10.1109/JIOT.2022.3143529.

[26] Xinran Zhang, Mugen Peng, Shi Yan, Yaohua Sun, Joint communication and computation resource allocation in fog-based vehicular networks, IEEE Internet Things J. 9 (15) (2022) 13195–13208, http://dx.doi.org/10.1109/JIOT.2022.3140811.

[27] Haijun Zhang, Lizhe Feng, Xiangnan Liu, Keping Long, George K. Karagiannidis, User scheduling and task offloading in multi-tier computing 6G vehicular network, IEEE J. Sel. Areas Commun. 41 (2) (2023) 446–456, http://dx.doi.org/10.1109/JSAC.2022.3227097.

[28] Kai Cui, Mustafa Burak Yilmaz, Anam Tahir, Anja Klein, Heinz Koeppl, Optimal offloading strategies for edge-computing via mean-field games and control, in: Proceedings of the 2022 IEEE Global Communications Conference, GLOBECOM, Rio de Janeiro, Brazil, 2022, pp. 976–981, http://dx.doi.org/10.1109/GLOBECOM48099.2022.10001412.

[29] Fabio Giust, Gianluca Verin, et al., MEC deployments in 4G and evolution towards 5G, in: ETSI White Paper No. 24, first ed., ISBN: 979-10-92620-18-4, 2018.

[30] Azzedine Boukerche, Robson E. De Grande, Vehicular cloud computing: Architectures, applications, and mobility, Comput. Netw. (ISSN: 1389-1286) 135 (2018) 171–189, http://dx.doi.org/10.1016/j.comnet.2018.01.004.

[31] Azzedine Boukerche, Robson E. De Grande, Vehicular cloud computing: Architectures, applications, and mobility, Comput. Netw. (ISSN: 1389-1286) 135 (2018) 171–189, 10.1016/j.comnet.2018.01.004.

[32] William Tärneberg, et al., Resource management challenges for the infinite cloud, in: 10th International Workshop on Feedback Computing at Cpsweek 2015, 2015.

[33] Shahin Vakilinia, Mustafa Mehmet Ali, Dongyu Qiu, Modeling of the resource allocation in cloud computing centers, Comput. Netw. 91 (2015) 453–470.

[34] Amitabha Ghosh, Andreas Maeder, Matthew Baker, Devaki Chandramouli, 5G evolution: A view on 5G cellular technology beyond 3GPP release 15, IEEE Access 7 (2019) 127639–127651, http://dx.doi.org/10.1109/ACCESS.2019.2939938.

[35] REM Maps, 5G-LENA, 2025, https://5g-lena.cttc.es/features/rem/, Retrieved 7 2025.

[36] 3GPP, 5G; Service Requirements for Enhanced V2X scenarios (3GPP TS 22.186 version 16.2.0 release 16), 2020, ETSI TS 122 186 V16.2.0, Nov. [Online]. Available: https://www.etsi.org/deliver/etsi_ts/122100_122199/122186/16.02.00_60/ts_122186v160200p.pdf.

[37] ETSI, Intelligent transport systems (ITS); vehicular communications; basic set of applications; part 2: Specification of cooperative awareness basic service, 2019, ETSI EN 302 637-2 V1.4.1, Apr. [Online]. Available: https://www.etsi.org/deliver/etsi_en/302600_302699/30263702/01.04.01_60/en_30263702v010401p.pdf.

[38] Dawei Wei, Junying Zhang, Mohammad Shojafar, Ning Kumari, Jianfeng Ma, Privacy-aware multiagent deep reinforcement learning for task offloading in VANET, IEEE Trans. Intell. Transp. Syst. 24 (11) (2023) 13108–13122, http://dx.doi.org/10.1109/TITS.2022.3202196.

[39] Alessandro Zanni, Se-Young Yu, Paolo Bellavista, Rami Langar, Stefano Secci, Automated selection of offloadable tasks for mobile computation offloading in edge computing, in: 2017 13th International Conference on Network and Service Management, CNSM, 2017, pp. 1–5, http://dx.doi.org/10.23919/CNSM.2017.8256026.

[40] GitLab Repository, 5G-LENA, 2024, https://gitlab.com/cttc-lena/nr/-/tree/master/examples, Retrieved 08 2024.

[41] NVIDIA, NVIDIA Tesla V100: The First Tensor Core GPU, https://www.nvidia.com/en-gb/data-center/tesla-v100/, Retrieved 11 2025.

[42] Spiceworks, What is WiFi 6? Meaning, speed, features, and benefits, 2022, https://www.spiceworks.com/tech/networking/articles/what-is-wifi-six/, Retrieved 7 2025.

[43] M. Chiappetta, AMD threadripper 3990X review: A 64-core multithreaded beast unleashed, 2020, https://hothardware.com/reviews/amd-ryzen-threadripper-3990x-cpu-review, February 7 . Retrieved 7 2025.

[44] M. Chiappetta, AMD Ryzen 9 3950X review: A 16-core zen 2 powerhouse, 2020, https://hothardware.com/reviews/amd-ryzen-9-3950x-zen-2-review, November 14 . Retrieved 7 2025.

[45] Wikipedia, Instructions per second, 2025, https://en.wikipedia.org/wiki/Instructions_per_second, Retrieved 7 2025.

[46] Kaneez Fizza, Nitin Auluck, Akramul Azim, Improving the schedulability of real-time tasks using fog computing, IEEE Trans. Serv. Comput. 15 (1) (2022) 372–385, http://dx.doi.org/10.1109/TSC.2019.2944360.

[47] Syed Sahil Abbas Zaidi, Mohammad Samar Ansari, Asra Aslam, Nadia Kanwal, Mamoona Asghar, Brian Lee, A survey of modern deep learning based object detection models, Digit. Signal Process. (ISSN: 1051-2004) 126 (2022) 103514, http://dx.doi.org/10.1016/j.dsp.2022.103514.

[48] G. de Préville, Paris: la vitesse moyenne d'un automobiliste en journée est de 13 1 km/h selon une étude, 2021, Le Figaro, Oct. 12 [Online]. Available: https://www.lefigaro.fr/societes/paris-la-vitesse-moyenne-d-un-automobiliste-en-journee-est-de-13-1-km-h-selon-une-etude-20211012, Retrieved 7 2025.

[49] CTTC, 5G-LENA, 2025, https://5g-lena.cttc.es/, Retrieved 7 2025.

[50] SUMO Project, SUMO, 2025, https://eclipse.dev/sumo/, Retrieved 7 2025.

[51] Manhattan-SUMO Documentation, Manhattan, 2025, https://sumo.dlr.de/docs/Tutorials/Manhattan.html, Retrieved 7 2025.

[52] Wendlasida Ouedraogo, Andrea Araldo, Badii Jouaber, Hind Castel, Remy Grunblatt, Can edge computing fulfill the requirements of automated vehicular services using 5G network? in: 2024 IEEE 99th Vehicular Technology Conference, VTC2024-Spring, Singapore, Singapore, 2024, pp. 1–5, http://dx.doi.org/10.1109/VTC2024-Spring62846.2024.10683522.

[53] Yeongjin Kim, Hyang-Won Lee, Song Chong, Mobile computation offloading for application throughput fairness and energy efficiency, IEEE Trans. Wirel. Commun. 18 (1) (2019) 3–19, http://dx.doi.org/10.1109/TWC.2018.2868679.

[54] Brian Benchoff, Pi 3 benchmarks: The marketing hype is true, 2016, https://hackaday.com/2016/03/01/pi-3-benchmarks-the-marketing-hype-is-true/, Retrieved 20 2024.

[55] William Zimmer, Raspberry pi 5: Les premiers benchmarks dévoilent des performances en hausse, 2023, https://www.phonandroid.com/raspberry-pi-5-les-premiers-benchmarks-devoilent-des-performances-en-hausse.html, Retrieved 7 2025.

[56] Versus, ARM cortex-a73 review: specs and price, 2024, https://versus.com/en/arm-cortex-a73, Retrieved 24 2024.

[57] Unicornplatform, How long should a GPU actually last? Plan for 3-5 years, 2023, https://unicornplatform.com/blog/how-long-should-a-gpu-actually-last-expect-3-5-years/, Retrieved 7 2025.

[58] Warehouse Technician, Warehouse technician salary in United States, 2025, https://www.indeed.com/career/warehouse-technician/salaries, Retrieved 7 2025.

[59] AWS, AWS lambda pricing, 2025, Available at: https://calculator.aws/#/addService/Lambda, . Retrieved 7 2025.

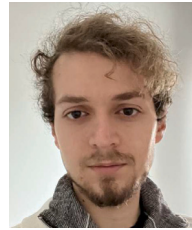

**Rosario Patanè** Ph.D. Student, University Paris-Saclay

Rosario Patanè is Ph.D. student at the University Paris-Saclay, in the Doctoral School of Sciences and Technologies of Information and Communication. He is pursuing his PhD in the specialty of Networks, Information, and Communication Sciences, with a focus on Vehicular Networks and Cloud/Edge Computing. His research explores the integration of Vehicular Cloud Computing into existing computational paradigms, aiming to assess the feasibility of replacing Edge Computing with Vehicular Computing. This to reduce infrastructure costs.

He holds a Master of Science in Computer Engineering from the University of Catania, where his thesis focused on energy-aware Proof of Work mechanisms for resource access management. Rosario also holds a Bachelor of Science in Computer Engineering from the same university. His bachelor thesis is about on the comparison of open-source home automation platforms.

In addition to his research, Rosario is involved in teaching at the University Paris-Saclay, delivering courses in Networking.

Rosario has published research on topics such as coinvestment in Edge Computing and the potential of Vehicular Cloud Computing to replace traditional Edge Computing in 5G networks. His work has been presented at conferences such as IEEE CCNC and IEEE WCNC.

His technical skills include various programming languages (Python, C++, Go, and others) and scientific softwares (NS3, SUMO, Omnet++). His native language is Italian, fluent in French, and English, with additional knowledge of Spanish and Latin.

Publications: - R. Patanè et al., "Coalitional Game-Theoretical Approach to Coinvestment with Application to Edge Computing", 2023 IEEE 20th Consumer Communications & Networking Conference (CCNC), Las Vegas, NV, USA. - R. Patanè et al., "Can Vehicular Cloud Replace Edge Computing?" 2024 IEEE Wireless Communications and Networking Conference (WCNC), Dubai, UAE.